\documentclass[12pt]{article}
\usepackage{graphicx}  
\begin{document}

\vspace{1cm}
\begin{center}
{\bf Food Web Structure and the Evolution of Ecological
Communities}\\

\vspace{0.5cm}

{\it Christopher Quince$^1$, Paul G. Higgs$^2$ and Alan J.
McKane$^1$}
\\
\bigskip
$^1$Department of Theoretical Physics and the $^2$School of
Biological Sciences,\\ University of Manchester, Manchester M13
9PL, UK \\
\end{center}

\section*{Abstract}

Simulations of the coevolution of many interacting species are
performed using the Webworld model. The model has a realistic set
of predator-prey equations that describe the population dynamics
of the species for any structure of the food web. The equations
account for competition between species for the same resources,
and for the diet choice of predators between alternative prey
according to an evolutionarily stable strategy. The set of species
present undergoes long-term evolution due to speciation and
extinction events. We summarize results obtained on the
macro-evolutionary dynamics of speciations and extinctions, and on
the statistical properties of the food webs that are generated by
the model. Simulations begin from small numbers of species and
build up to larger webs with relatively constant species number on
average. The rate of origination and extinction of species are
relatively high, but remain roughly balanced throughout the
simulations. When a `parent' species undergoes speciation, the
`child' species usually adds to the same trophic level as the
parent. The chance of the child species surviving is significantly
higher if the parent is on the second or third trophic level than
if it is on the first level, most likely due to a wider choice of
possible prey for species on higher levels. Addition of a new
species sometimes causes extinction of existing species. The
parent species has a high probability of extinction because it has
strong competition with the new species. Non-parental competitors
of the new species also have a significantly higher extinction
probability than average, as do prey of the new species. Predators
of the new species are less likely than average to become extinct.

\section{Introduction}
\label{intro}

In this article we discuss the Webworld model which, by describing the 
interaction of species over a wide range of timescales, allows us to 
start from a very few species and evolve food webs which represent the 
predator-prey relationships amongst a large community of diverse 
species. A typical food web generated by the model is shown in Figure 1. 

\begin{figure}[h]
\begin{center}
\includegraphics[width=6.0cm,trim = 0 160 0 0,clip = 
true]{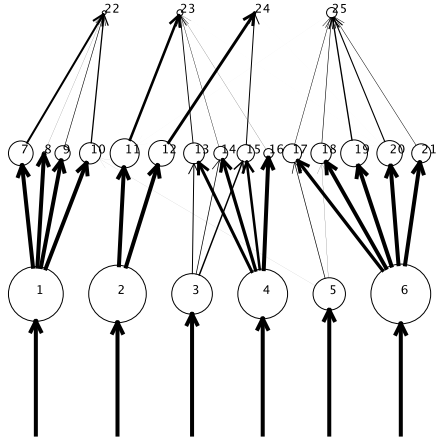}
\end{center}
\caption{\em Typical food web produced by the model} \label{initial}
\end{figure}

The web shown in the figure is typical in the sense that the webs which
are evolved by Webworld resemble real food webs by having very few 
predator-prey links between species on the same level, and also very few 
cycles; most of the links, and certainly the major ones, start on one level 
and end on a higher one. A more quantitiative comparison between the model 
and ecological data confirms this broad similarity.

Webworld was designed as a model of an evolving community of many
interacting species. It is intended to address questions of
interest on both ecological and evolutionary timescales. Models of
population dynamics accounting for predator-prey interactions
between species work on the ecological scale. These date back to
Lotka and Volterra and form a familiar part of the theoretical
biology literature (Pielou, 1977; Roughgarden, 1979). Many studies 
have considered the dynamical behaviour and type of attractors that 
arise with systems of coupled equations representing a few (usually 
two or three) species (Emlen, 1984; Hallam, 1986; Hastings \& Powell,
1991; McCann \& Yodzis, 1994; Post {\it et al.} 2000). Some studies 
have addressed the problem of stability of these dynamical equations 
when many interacting species are present (May, 1974; Svirezhev 
\& Logofet, 1983; Logofet, 1993; Hofbauer \& Sigmund, 1998). When 
one considers many species, the structure of the food web that connects 
them becomes relevant. Which species prey on which other species? 
Which species compete with one another for the same resources? Recent 
studies 
of general food web structures including these features have
been carried out by Bastolla {\it et al.} (2000) and L\"assig {\it et 
al.}
(2001).

There is also a body of
literature in theoretical biology that considers the statistical
properties of food webs (Pimm, 1982), both as observed in the
field (e.g. Hall \& Raffaelli, 1991; Goldwasser \& Roughgarden,
1993; Martinez \& Lawton, 1995) and as predicted by various types
of random graph models (Cohen, 1990; Cohen {\it et al.} 1990).
These studies consider how many trophic levels there are in food
webs, how the species are distributed between the levels, and how many
predators and prey are possessed by each species in the
web. Patterns such as these are static rather than dynamic, but it
is clear that the arrangement of the links in the food web will
influence the population dynamics, and that the community
structure we observe has been created by the dynamics of the
system, rather than being simply thrown together. We therefore
require theoretical models that can address both static and
dynamic questions.

In addition to the relatively rapid ecological dynamics, it is
important to consider long-term evolutionary dynamics. Speciation
will create new species and extinction will remove some of the old
ones. When a new species arises, this will have knock-on effects
on other species in the community through predation and
competition interactions. There is the potential for indirect
effects to influence many other species. There are now a large
number of evolutionary models, mostly in the physics literature,
that investigate the dynamics of extinction events and the
possibility of large scale avalanches of extinctions. Mass
extinctions are an important feature observed in the fossil
record. Biologists have tended to ask what causes these 
events --- climate changes, meteorite strikes etc. An idea stemming 
from the theory of self-organized criticality is that the dynamics 
of the system itself may be inherently unstable and subject to large
scale avalanches of extinctions (Bak \& Sneppen, 1993; Sol\'e {\it
et al.} 1997). Theoretical models of macro-evolution and
extinction have been reviewed recently by Drossel (2001). It is
not surprising that changes in the non-living environment of an
ecosystem can sometimes have catastrophic effects on the living
species. The key question here is what would the
macro-evolutionary dynamics be like in the absence of external
changes. Does evolution lead to communities of stable interacting
species that cannot be replaced by new ones? Does it create a
continual turn-over of new species replacing old ones (a Red Queen
scenario), or does it create critical food web structures prone to
large scale fluctuations?

Webworld has been designed to fill a gap between the different
types of models described above. Neither the models of food web
structure or those of population dynamics consider evolution,
whilst the macro-evolutionary models contain very little
biological detail and often ignore population dynamics. The model
we have arrived at involves lengthy computer simulations and is
more complex than many of the other models referred to above.
Whilst we recognize that simplicity is a virtue, we also feel that
the level of detail included here has the payoff of allowing us to
address a very wide range of biological questions. We also would
like to resist the tendency to assume that effects observed in
very simple models will always prove to be `universal'. We will
argue that it is important to have qualitatively correct population 
dynamics, and point out cases where omitting seemingly unimportant 
effects, leads to significant changes in the biological conclusions.

The Webworld model has already been described in detail in Drossel
{\it et al.} (2001) and Caldarelli {\it et al.} (1998), therefore
we only describe it briefly in the following section. This paper
will try to summarize some of the effects observed in simulations
with Webworld in our previous papers, and will then focus on the
question of what happens when a new species is added to a food
web. Diagrams will be shown of a few representative cases, and new
statistical results will be given describing the response of the
web to a single speciation event.

\section{Population Dynamics}
\label{model}

Each species in the model is represented by a set of $L$ features
or phenotypic characters chosen from a set of $K$ possible
features. Typically $L = 10$ and $K = 500$ in the model, which
means that the number of possible species is extremely large
and evolution never runs out of scope for innovation. Each species
has a ``score" against any other species that is calculated as a
function of the set of features possessed by each of the species.
The score $S_{ij}$ is positive if species $i$ is adapted to prey
on species $j$, and is zero if not. These scores are used as the
numerical coefficients in the population dynamics equations
discussed below.

Let the rate at which one individual of species $i$ consumes
individuals of species $j$ be denoted by $g_{ij}(t)$. This is
usually called the `functional response', and it depends in
general on the population sizes. We suppose that the population
size $N_i$ of each species satisfies an equation of the form:
\begin{equation}
\frac{dN_i(t)}{dt} = -N_i(t)+ \lambda \sum_{j}N_i g_{ij}(t) -
\sum_j N_j g_{ji}(t).
\label{popsize}
\end{equation}
The first term on the right represents a constant rate of death of
individuals in absence of interaction with other species. The
final term is the sum of the rates of predation on species $i$ by
all other species, and the middle term is the rate of increase of
species $i$ due to predation on other species. Where there is no
predator-prey relationship between the species the corresponding
rate $g_{ij}$ is zero. The factor $\lambda$ is less than 1, and is
known as the ecological efficiency.  It represents the fraction of
the resources of the prey that are converted into resources of the
predator at each stage of the food chain. Throughout this paper,
we have taken $\lambda=0.1$, a value accepted by many ecologists
(Pimm, 1982).

The external environment is treated as an additional `species 0'.
For primary producers, the middle term includes a non-zero rate
$g_{i0}$ of feeding on the external resources. External resources
(e.g. sunlight) enter the ecosystem at a constant rate $R$. In the
equations this is implemented by defining $N_0 = R/\lambda$, and 
keeping $N_0$ fixed. We have deliberately chosen the form of
Eq.~(\ref{popsize}) to be the same for all species. We do not want
to define different equations for primary producers, herbivores,
and carnivores etc, because species can change their position in
the ecosystem as it evolves, and most species are both predators
and prey.

The most straightforward form for the functional response would be
to have $g_{ij}$ proportional to the prey population size $N_j$,
as is the case in the Lotka-Volterra equations (Pielou, 1977;
Roughgarden, 1979). A variety of other forms have been
proposed that account for the fact that when prey are scarce or
when many predators choose the same prey, competition between
predators reduces the amount of prey available to each predator,
whilst when prey are abundant, the consumption rate per predator
must saturate rather than continue to increase indefinitely with
the prey population size (Holling, 1959; Beddington, 1975; 
Huisman \& De Boer, 1997). The form of the functional response used
in the recent versions of Webworld is

\begin{equation}
g_{ij}(t) = \frac{S_{ij}f_{ij}(t)N_j(t)}{bN_j(t) +\sum_k
\alpha_{ki}S_{kj}f_{kj}(t)N_k(t)}.
\label{gij}
\end{equation}
This is based on the ratio-dependent functional responses used by
Arditi \& Ginsburg (1989) and Arditi \& Michalski (1995). We have
described in detail in Drossel {\it et al.} (2001) how we generalized 
these studies to give equation (\ref{gij}). In the denominator,
the sum runs over all the species that are predators of species
$j$. The factor $\alpha_{ki}$ determines the strength of
competition between species for the same resources. This depends
on the degree of similarity between the species:
\begin{equation}
\alpha_{ij}=c+(1-c)q_{ij}, \label{overlap}
\end{equation}
where $c$ is a constant such that $0 \le c <1$, and where $q_{ij}$
is the `overlap', or fraction of features of species $i$ that are
also possessed by species $j$. This means that competition is
strongest between similar species (or members of the same
species), and is weaker for different species because they can use
the resources in slightly different ways.

The factor $f_{ij}$ is the fraction of its effort (or available
searching time) that species $i$ puts into preying on species $j$.
These efforts must satisfy $\sum_j f_{ij}=1$. We suppose that the
efforts of any species $i$ are chosen so that the gain per unit
effort $g_{ij}/f_{ij}$ is equal for all prey $j$. If this were not
true, the predator could increase its energy intake by putting
more effort into a prey with higher gain per unit effort. This
choice of efforts leads to the condition

\begin{equation}
f_{ij}(t) = \frac{g_{ij}(t)}{\sum_k g_{ik}(t)}.\label{eff}
\end{equation}

We showed (Drossel {\it et al.} 2001) that this choice is an
evolutionarily stable strategy (ESS) (Parker \& Maynard Smith,
1990; Reeve \& Dugatkin, 1998). If the population has efforts
chosen in this way, there is no other strategy with a different 
choice of efforts that can invade the population. Predator diet
choice is often discussed using optimal foraging theory (Stephens and
Krebs, 1986). The basic idea is that each predator maximises its individual
rate of resource input. In our model, the success of the predators' strategies
is a function of predator population size and the strategies of the other
predators. The ESS solution is not equivalent to maximizing the resource input
for a single predator. In most other models of optimal foraging, there is no
competition between predators, hence there is no distinction between the ESS and the
strategy with maximal resource input rate.  

There are only four principal parameters that determine the
behaviour of the model: $R$, the rate of input of external
resources; $\lambda$, the ecological efficiency; $b$, which
controls the saturation level of the functional response; and $c$,
which controls the competition strength. All other quantities are
generated automatically in the simulation. For example, the scores
$S_{ij}$ and the values $\alpha_{ij}$ are determined by the
features of the species. The efforts are also determined at each
time point in the dynamics by ensuring that they remain at their
ESS value (see Drossel {\it et al.} 2001 for more details).

\section{Evolutionary Dynamics}
\label{evol}

For any set of species in the web the population dynamics can be
described using the above equations. Evolution in Webworld occurs
by speciation events. An existing species is chosen at random to
undergo speciation, and a new species is created that differs by
one randomly chosen feature from the parent species. The
population dynamics is then followed with the additional species
until a new stable state is reached. Species whose population
falls below a threshold of 1.0 are considered extinct and removed
from the web. In this way, new species gradually replace older
ones, and the number of species present rises and falls.

It is worth stressing that the features should not be thought of 
as having any genetic basis --- they are purely phenotypic 
characteristics --- nor should the random replacement of randomly 
chosen features be thought of as genetic mutations. We do not attempt 
to model the extremely complex process of speciation. Instead, we
imagine observing the variation of the species in the system on 
such a coarse time scale that the process of speciation appears 
to be stochastic. This is the process that we are modelling when we 
randomly change features. Of course, there are several variants 
of the scheme which we have adopted which it could be argued might 
have done just as well. For instance, the parent could be chosen 
according to criteria dependent on population size or trophic level, 
and not purely at random, and the child 
might not always have just one modified feature. We hope in the
future to check that changes such as these do not alter the 
general predictions of the model. 

\begin{figure}
\begin{center}
\rotatebox{270}{\scalebox{1.8}{\includegraphics[width=.3\textwidth]{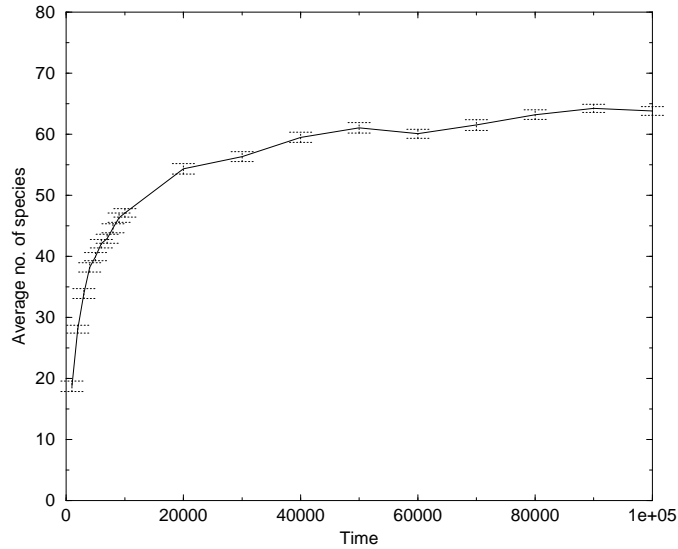}}}
\end{center}
\caption[]{Mean number of species vs. time} \label{fig1}
\end{figure}

Results in this paper are obtained from 80 different runs of
Webworld using the same set of parameters ($\lambda$ = 0.1, $c$ =
0.5, $b$ = 0.005, and $R = 1.0 \times 10^{5}$). Each run lasted for 
100000 speciation events. The mean number of species in these 80 runs 
is shown as a function of time in Figure 2. This shows a fairly rapid
increase initially, followed by stabilization to a fairly constant
state where speciations are balanced by extinctions. Our previous
paper (Drossel {\it et al.} 2001) shows the variation in the
number of species in several individual runs. There is
considerable fluctuation in each individual run, even at the later
times when the average number is relatively constant. This shows
that there is a continual turnover of species. There is
considerable fluctuation between different runs with the same
parameters due to the random choice of the feature that is changed
at each speciation event.

Figures 3-6 show some examples of particular food webs generated
by these simulations. In order to indicate population sizes,
circles are shown with radius proportional to the log of the
population size. Typical population sizes decrease by an order of
magnitude on each successive trophic level. The arrows represent
the flow of resources from prey to predator. The thickness of the
arrow represents the amount of effort the predator puts into each
prey. Many of the second and third level species split their
effort between more than one prey. The level assigned to each
species is determined by the shortest path to the external
resources. For example, species 20 in Figure 3 is on level 2
because it feeds on species 6, which feeds on external resources.
There are also longer pathways from species 20, via species 23 and
24. We use the shortest path to define the level because this can
always be calculated unambiguously (even in the presence of
cycles) and because longer pathways provide relatively little
resources. One observation from these webs is that the majority of
interactions are simply between a prey species on one level and a 
predator
on the level above (an `upward' arrow in the diagrams).
 More complex situations, such as `downward' or
`horizontal' arrows occur rarely.

In Figure 4, species 1-25 were in a state with stable populations.
Species 1 has just undergone speciation to create the new species
26, which is also a level 1 species, like its parent. Figure 4
shows the situation after the population sizes have reached a new
equilibrium. No extinction events have occurred. The population of
species 26 has risen to be comparable to that of species 1. There
is increased competition on level 1, hence some of the level 1
species populations have decreased (e.g. species 5). In Figure 3,
species 1 had four predators feeding almost exclusively on it.
When species 26 rises to a large population size, it pays these
predators to put some of their effort into the new prey. This can
be seen in Figure 4, where species 8 and 9 have switched
predominantly to feeding on species 26, whilst species 7 has
species 26 as a minor food source.

Another example is shown in Figures 5 and 6. Here we begin with the 
same
25 initial species, and species 26 arises on level 2, as a result
of speciation of species 11. This new species causes the
extinction of its parent species, 11, and three other species, 4,
16 and 24. The species that go extinct are all shaded in Figure 5.
Figure 6 shows the situation after the new stable state is
reached. This example shows that extinctions can occur on all
levels, and that the nature of the interaction between the new
species and the species that become extinct is not always
straightforward. This point is pursued in the following section.

\newpage
\twocolumn
\begin{figure}[t]
\begin{center}
\includegraphics[width=6.0cm,trim = 0 160 0 0,clip = 
true]{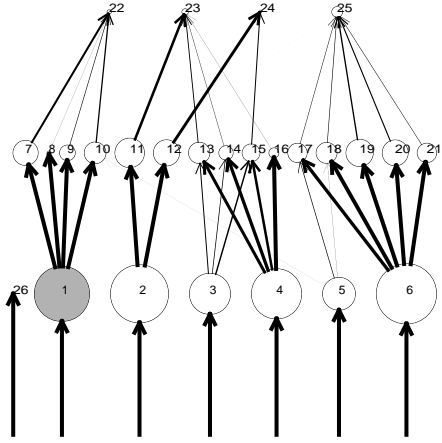}
\end{center}
\caption{\em Stable addition event: iteration 0} \label{sta1i0}
\end{figure}

\begin{figure}[b]
\begin{center}
\includegraphics[width=6.0cm,trim = 0 160 0 0,clip = 
true]{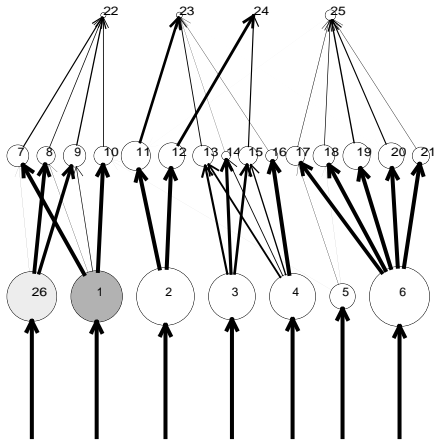}
\end{center}
\caption{\em Stable addition event: equilibrium} \label{sta1i10001}
\end{figure}

\begin{figure}[t]
\begin{center}
\includegraphics[width=6.0cm,trim = 0 160 0 0,clip = 
true]{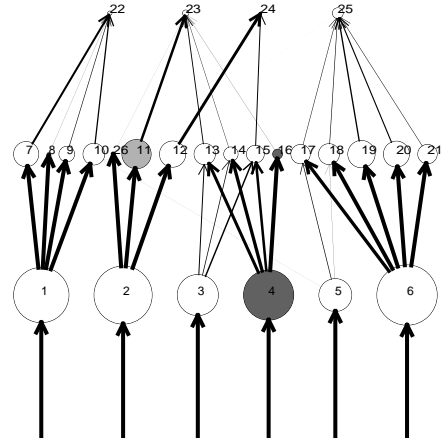}
\end{center}
\caption{\em Deletion event: iteration 0} \label{del11i0}
\end{figure}

\begin{figure}[b]
\begin{center}
\includegraphics[width=6.0cm,trim = 0 160 0 0,clip = 
true]{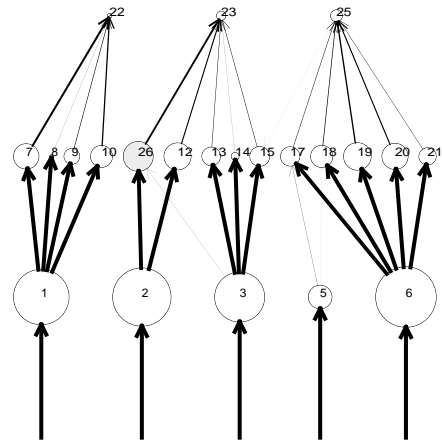}
\end{center}
\caption{\em Deletion event: equilibrium} \label{del11i10001}
\end{figure}

\onecolumn
\newpage

We referred to the continual turnover of species that occurs with
this model. The main reason for this is that coevolutionary
effects act both upwards and downwards in the web. New species on
level one, for example, can out-compete older level 1 species,
leading to their extinction, and possibly to the extinction of any
level two species that fed on the old species (an upwards effect).
However, a new level 2 species could arise that is a better
adapted predator that causes its level 1 prey to go extinct (a
downward effect). The way these effects operate depends on the
form of the population dynamics equations. In our original paper
(Caldarelli {\it et al.} 1998) we used a much simpler form of the
population dynamics in which downward effects did not occur. We
observed that the rate of turnover of species became slower and
slower, and that the chance of a newly added species surviving
decreased to virtually nil. Without downward effects, level one
species can arise that are increasingly better adapted to the
external environment, and it becomes increasingly more difficult
to improve on them. With coevolutionary effects in both
directions, species that are successful at one time do not remain
successful for ever.

\begin{figure}
\begin{center}
\rotatebox{270}{\scalebox{1.8}{\includegraphics[width=.3\textwidth]{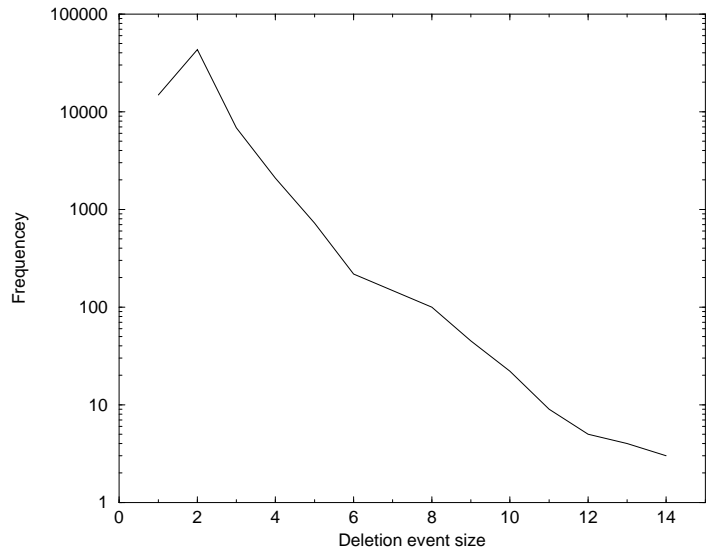}}}
\end{center}
\caption[]{Deletion event size distribution} \label{fig6}
\end{figure}

The sensitivity that we observed of the evolutionary dynamics to
the short time scale population dynamics makes us somewhat
sceptical of the results of some of the simpler macro-evolutionary
models in which there is essentially no population dynamics
included at all. One question that has frequently been studied is
the distribution of sizes of extinction events. Figure 7 shows the
distribution of the number of species going extinct in each speciation 
event,
given that at least one existing species goes extinct.
The parameters are as for the runs in Figure 8.
 The majority of events are
small, although the largest events (up to 14 species) represent a
considerable fraction of the total number of species (average of
59 species per web). If we wished to look for evidence of critical
phenomena it would be natural to ask how the size of the largest
extinction events depends on the size of the web. However, in our
model, the web size is not an independent parameter. We can
increase the number of species by increasing the external
resources, $R$, or decreasing the competition strength, $c$ (see
examples in Drossel {\it et al.} 2001). Computer time increases
rapidly with the number of species, however, and it would be
impractical to consider webs that were an order of magnitude
larger than the present ones. 

\section{Food Web Properties and Co-evolution}

For the same 80 runs discussed above, details of the web were
stored at intervals of 10000 speciation events from time 10000 to
100000, thus giving 800 largely independent webs which we
investigated in detail. Table 1 shows average properties of these
webs. In this analysis a link was defined as being present if the
effort of the predator against the prey was greater than 1\%.
Since the efforts are proportional to the consumption rates
(because of the ESS criterion), this means that a link is counted
if greater than 1\% of the resources consumed by the predator come
from the prey. This is slightly different to the definition used
in our previous papers, which required $g_{ij}$ to be greater than
1.0 for a link to be counted. The figures in Table 1 can be
compared with those in Drossel {\it et al.} (2001), which use a
range of different parameters, and those in Caldarelli {\it et
al.} (1998), which were generated using a simpler form for the
population dynamics equations.

In all the webs contributing to Table 1, the maximum trophic level
present is either 3 or 4, with a mean of 3.85. The average level
of all species is 2.31. The distribution of species between levels
is given in Table 2. With these parameters, level 4 species tend
to have very low populations because there are barely sufficient
resources coming up through the web to maintain them. Species can
also be classed as basal (having no prey), top (having no
predators) or intermediate (having both predators and prey). A
majority of species are intermediate for most reasonable parameter
values. In fact the 10\% of top species observed here is rather higher 
than
the values in the examples in our previous papers. This is due to the 
rule used to define a link. Very weak interactions that would have been 
counted with our previous definition are discounted here using the 
criterion that the effort must be greater than 1\%. We will consider in 
more
detail elsewhere the way web statistics depend on the link definition.
 Also shown in Table 1 are the mean overlaps between species on each
level (see equation (\ref{overlap})). This shows that considerable
phenotypic diversity has accumulated in the species: only 2.6 out
of 10 features are shared between species on level 1, and less
than this on the higher levels.

The general patterns observed in real food webs are reproduced
fairly well by the model. The values of quantities in real webs
fluctuate greatly from web to web (see examples in Caldarelli 
{\it et al.} 1998). This can be put down to three factors. Firstly, 
the random nature of the evolutionary process leads to quite large
fluctuations in web properties even if physical properties such as
resource input and ecological efficiency are the same. In the
model, this can be seen from the standard deviations in the web
properties. Secondly, real webs are observed in different types of
locations (lakes, deserts, estuaries etc.) that clearly do differ
in terms of resource input and the nature of the limiting
resource. Different groups of organisms exist in different
locations that may differ in their ecological efficiency. In terms of 
the
model, we
began to consider the way the web properties changed with
parameters such as $R$, $\lambda$ and $c$ in our previous papers. 
The third source of variation between the real webs is due to human
observation. Different researchers use different definitions of
what counts as a species, and what counts as a link. To what
extent should similar species be lumped together? How often must a
predation event be observed before a link is defined as being
present? These questions are not straightforward, and they make
detailed comparison between individual real webs, and between real
and model webs difficult. Furthermore, while it is the case that the 
webs that we have generated are being compared to local communities, 
be they islands, lakes or forests, it is clear that the interaction 
between the community that we have evolved, and the species and 
individuals 
outside it, needs to be included in order to make the comparison with 
data
valid. We hope to include the effects of immigration, and other 
such factors, in future work.

\begin{table}
\caption{Statistics on Food Web Structure}
\begin{center}
\renewcommand{\arraystretch}{1.4}
\setlength\tabcolsep{5pt}
\begin{tabular}{llllll}
\hline\noalign{\smallskip} Result&Mean&Error in mean&Std. dev.\\
\noalign{\smallskip} \hline \noalign{\smallskip} No. of
species&59.1&0.3&8.6\\ Links per species&1.694&0.005&0.128\\ Av.
level&2.313&0.002&0.059\\ Av. max. level&3.85&0.01&0.36\\ Basal
species(\%)&12.44&0.06&1.80\\ Intermediate
species(\%)&77.48&0.18&4.92\\ Top species(\%)&10.06&0.16&4.45\\
Mean overlap level 1&0.260&0.002&0.047\\ Mean overlap level
2&0.105&0.001&0.019\\ Mean Overlap level 3&0.089&0.001&0.026\\
\hline
\end{tabular}
\end{center}
\label{Tab1}
\end{table}

\begin{table}
\caption{Distribution of Species between Trophic Levels}
\begin{center}
\renewcommand{\arraystretch}{1.4}
\setlength\tabcolsep{5pt}
\begin{tabular}{llllll}
\hline\noalign{\smallskip} Level&Number of Species&Std.
dev.&Proportion of Species&Std. dev.\\ \noalign{\smallskip}
\hline \noalign{\smallskip} 1&7.30&1.14&0.125&0.018\\
2&27.43&5.21&0.462&0.042\\ 3&22.97&4.29&0.389&0.046\\
4&1.64&0.70&0.029&0.014\\ \hline
\end{tabular}
\end{center}
\label{Tab2}
\end{table}

We have used the set of 800 stored webs to study in detail what
happens when a single speciation event occurs. We performed 2000
independent speciation events on each of the starting webs and
recorded the properties of each of the webs that arose when a new
equilibrium situation was reached. Table 3 shows the statistics of
these events. Events were defined as either addition (the child
species survives and nothing goes extinct), substitution (the
child species replaces the parent species but no other extinctions
occur), no change (the child species goes extinct immediately,
leaving the web the same as before the speciation event), and
deletion (anything that is not addition, substitution or no
change). In a deletion event, the new species must cause at least
one species to become extinct that is not its parent. It can be
seen approximately 89\% of attempted speciations lead to no
change, and the other 11\% lead to a change in the web of some
description.

We also show in Table 3 the origination rate (i.e. the probability
that the newly created species survives), and the extinction rate
(the average number of species excluding the new species that go
extinct per speciation event). These quantities are both
approximately 8.9\%, and do not differ significantly from each
other. In other words, originations balance extinctions on the
timescale of a single speciation event. However Figure 2 shows
that, on a long timescale, there is a slight increase in the total
number of species present over the period 10000 to 100000
speciation events during which the stored webs were collected.
Thus there is a very slight excess of originations to extinctions.
Figure 8 shows the way the extinction and origination rate vary
with time. In the initial part of the simulation up to time 10000,
both rates are high and are decreasing significantly with time.
For the period 10000 - 100000 used for the statistics, the two
rates are relatively constant. They seem to leveling off at roughly
8\%, which represents a moderate non-zero rate of turnover of
species.

\begin{figure}
\begin{center}
\rotatebox{270}{\scalebox{1.8}{\includegraphics[width=.3\textwidth]{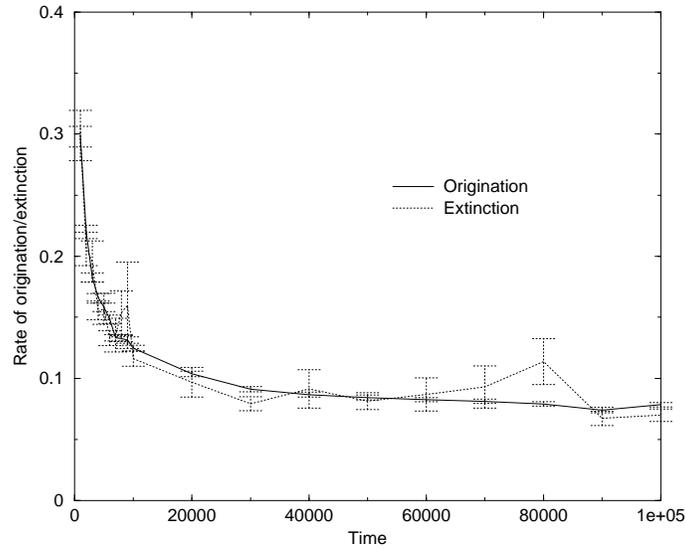}}}
\end{center}
\caption[]{Rate of origination/extinction vs. time} \label{fig7}
\end{figure}

The standard deviations shown in Table 3 measure variations
between the 800 stored webs. The quantities were averaged over the
2000 speciation events sampled for each stored web, and the mean and 
standard
deviations were then taken over the 800 webs. For most quantities
the standard deviation is small compared to the mean, but this is
not true for the deletion probability and the extinction rate.
When the distribution of extinction rates is considered across
webs, there is a significant tail of high extinction rate webs
that causes the large standard deviation. In contrast, the
origination rate varies much less between webs. We intend to
consider in more detail in future what properties of the stored
webs make them more or less prone to extinctions.

\begin{table}
\caption{Probabilities of event types arising from a single
speciation}
\begin{center}
\renewcommand{\arraystretch}{1.4}
\setlength\tabcolsep{5pt}
\begin{tabular}{llllll}
\hline\noalign{\smallskip} Event type&Mean&Error in mean&Std.
dev.\\ \noalign{\smallskip} \hline \noalign{\smallskip}
Addition&0.0417&0.0005&0.0152\\
Substitution&0.0230&0.0003&0.0093\\
Deletion&0.0425&0.0035&0.1000\\ No change&0.8928&0.0033&0.0932\\
\noalign{\smallskip} \hline \noalign{\smallskip} Origination
rate&0.08853&0.0008&0.0231\\ Extinction
rate&0.08949&0.0038&0.1075\\ \hline
\end{tabular}
\end{center}
\label{Tab3}
\end{table}

Table 4 considers the way the trophic level of the parent species
undergoing speciation affects the process of addition of new
species. We have defined the level-specific origination rate as
the probability that the child species survives given that the
parent was on a particular level. This varies substantially
between levels. It can be seen that the figure for level 2 is
roughly equal to the average rate for the whole web given in Table
3, whereas the origination rate is significantly higher than
average for level 3 and significantly lower for levels 1 and 4.
Table 4 also gives the probability that the child ends up in each
level given that it survives and given the level of the parent.
Most of the child species occupy the same level as the parent
(e.g. 92\% of the children of level 1 species are also in level
1). There is a relatively small amount of movement between levels.
The largest of the off-diagonal terms is the 29\% probability of
the child being in level 3 if the parent is in level 4. Given that most
species enter the web on the same level as their parent, we can
interpret the level-specific origination rates as measuring how
easy it is to add species in a given level. The rate in level 1 is
low because competition for external resources is strong, and
because there is only one type of resource to compete for. For
species on levels 2 and 3 there are many prey species on the level
below from which to choose, and it becomes easier to find a niche
in which to survive. The origination rate in level 4 is low, and
we believe this is because level 4 species have a difficult time
surviving due to lack of resources. Even a small amount of
competition on this level is sufficient to drive the population
below the minimum extinction threshold of 1. We have also measured
the proportions of the species on each level as a function of
time. These are roughly constant over the period 10000 - 100000,
indicating that additions and deletions are roughly equal on each
individual level.

\begin{table}
\caption{The relationship of the trophic level of the child
species to that of the parent, and the dependency of the
origination rate on the parental level.}
\begin{center}
\renewcommand{\arraystretch}{1.4}
\setlength\tabcolsep{5pt}
\begin{tabular}{lllllrll}
\hline\noalign{\smallskip} Child level:& 1 & 2 & 3 & 4 &
Origination rate& Error & Std. dev.\\ \noalign{\smallskip} \hline
\noalign{\smallskip} Parent level& & & & & & & \\
1&0.92&0.00&0.02&0.06&0.0200&0.0005&0.0139\\
2&0.00&0.90&0.09&0.02&0.0893&0.0009&0.0256\\
3&0.00&0.01&0.97&0.02&0.1117&0.0012&0.0334\\
4&0.00&0.03&0.29&0.68&0.0638&0.0025&0.0642\\ \hline
\end{tabular}
\end{center}
\label{Tab4}
\end{table}

Table 5 considers cases where a newly-created species survives,
and asks what is the relationship between this species and any
other species that go extinct as a consequence. The mean number of
species affected is the mean number of species going extinct in
each category as the result of the origination of one species. The
mean probability of extinction is the probability that a species
in a given category becomes extinct. The row `all species' refers
to all species previously in the web before the speciation event.
`Parent' refers to the parent of the new species. The `Predators'
and `Prey' rows consider only species that are predators or prey
of the new species. Note that the efforts of each species change
during the population dynamics (see discussion in Drossel {\it et
al.} 2001), so that predation links in the food web switch on and
off as the population sizes change. For the purposes of this
table, a species counts as a predator or prey if a link between
the species is present at any time during the population dynamics
(i.e. if the appropriate effort become greater than 1\% for at
least some of the time). Competitors are species that share at
least one prey species with the newly-added species for at least
some of the time during the population dynamics. Non-parental
competitors are competitors other than the parent of the new
species. Competitors' predators are species that are predators of
the competitors of the new species.

From the top row, we see that an average of 0.773 species of all
types become extinct per origination. This corresponds to a
probability of only 1.4\% that any random species goes extinct. In
comparison, there is a 39.2\% probability that the parental
species goes extinct. This is due to the strong competition
between parent and child. Since they differ by only one feature
out of 10, the overlap value will be 0.9, and hence the $\alpha$
value will be very high. Typical overlaps between species on a
level are between 0.089 and 0.26 (from Table 1), hence competition
between parent and child species is much stronger than for most
species pairs. Also, since they share nearly all features, the two
species will often be adapted to feed on the same prey, which
again increases competition above that typical for two species on
the same level. For non-parental competitors, the extinction
probability is 3.4\%, which is about two and a half times higher
than the average. Note that the statistical errors in all the
quantities in Table 5 are small and are therefore not given. Thus,
competition with the new species is a significant factor in
causing extinction of existing species.

The extinction probability for prey of the new species is also
significantly higher than average, indicating that new,
well-adapted predators can drive their prey extinct. However this
value is less than the value for non-parental competitors. In
contrast, the extinction probability for predators of the new
species is significantly less than average. This is to be
expected, because a predator of a successful new species is also
likely to be successful. The final case of competitors' predators
was considered because one might expect that if competitors of the
new species are driven extinct then predators of those species
would also be more likely to go extinct than average. The result
shows that the competitors' predators extinction rate is equal to
the average rate, hence the effect we looked for is not apparent.
One reason for this is that there are a rather large number of
competitors' predators, hence the figure is bound to be rather
close to the average. Also, species on the upper levels tend to
have several prey, so that extinction of any one of these is not
particularly important.

\begin{table}
\caption{The relationship of species going extinct to the newly
added species.}
\begin{center}
\renewcommand{\arraystretch}{1.4}
\setlength\tabcolsep{5pt}
\begin{tabular}{lll}
\hline\noalign{\smallskip} Trophic relationship&Mean Number of
Species Affected &Mean Probability of Extinction\\
\noalign{\smallskip} \hline \noalign{\smallskip} All species
&0.773&0.014\\ Parent&0.392&0.392\\ Predators&0.017&0.008\\
Prey&0.073&0.021\\ Competitors&0.499&0.147\\ Non-parental
Competitors&0.108&0.034\\ Competitors' predators&0.082&0.014\\
\hline
\end{tabular}
\end{center}
\label{Tab5}
\end{table}

\section{Discussion}

The most distinctive feature of Webworld, and the aspect which sets it 
apart 
from other models in this field, is the attempt to model phenomena 
which 
occur on very different time scales. On the shortest time scale in the 
model, 
the number and types of species are fixed, as are the number of 
individuals
that belong to these species. Only the amount of effort individuals of 
species $i$ put into preying on individuals of species $j$, $f_{ij}$, 
is 
allowed to vary. The choice of $f_{ij}$ is a choice of foraging 
strategy, in 
our case given by the self-consistent solution of equations (\ref{gij}) 
and
(\ref{eff}) (with the $N_i$ fixed). On longer time scales the number 
and type 
of species is still fixed, but the number of individuals of a given 
species 
is now allowed to vary. This is the realm of traditional population 
biology,
and in Webworld corresponds to determining the solutions of equation
(\ref{popsize}) in the long time limit. For simplicity, we have taken 
the
death rate to be the same for all species. The choice of making the
death rate equal to unity in equation (\ref{popsize}) sets the 
timescale 
for the population dynamics. In all the simulations which we have 
carried out, 
we find that these solutions are steady states; no limit cycles or 
chaotic 
behaviour have been observed. 

This point is worthy of further comment. A large fraction of the 
research 
which has been carried out in theoretical population dynamics has been 
concerned with models involving two or three species. There has been 
comparatively little work carried out on generic multispecies 
communities 
where a typical species will have several predators and several prey
(whose number and identity may change with time) which is the situation 
of 
interest to us here. As a consequence, attention has focussed on the 
relatively simple equations found when only a few species are present, 
and
especially on the phenomena of limit cycles and chaos frequently found 
in
such equations. It is an open question as to whether these effects will 
be 
seen in food webs with a large number of species. One can argue that if 
one 
species is coupled to a large number of other species, these will act 
as a 
reservoir and blur out the details of the interactions that cause 
chaos, 
and so lead to a simpler dynamics. Alternatively, the adaptive nature 
of 
some aspects of the dynamics may lead to configurations where such 
behavior 
is less likely. This argument is similar to that put forward by 
Berryman \& Millstein (1989a,b), who suggested that natural selection 
might 
favour parameter values which minimize the likelihood of chaotic 
dynamics. 
Of course, there may be other reasons why we have only seen steady 
states. 
Since the study of the long-time dynamics of equation (\ref{popsize}) 
was 
not the ultimate goal of this investigation, we did not carry out a 
comprehensive investigation in the entire space of possible solutions, 
and 
it may be that some more complex behaviour does in fact exist. It might 
also be that the form of our equations has a particular structure which 
precludes more complicated long-time behaviour. Further work is 
required 
to decide which, if any, of these explanations is the correct one.

On the third, and longest, time scale the number and type of species is 
allowed to vary through a speciation mechanism. Effects of this kind 
have 
been modelled far less than have population dynamics or foraging 
mechanisms.
One reason is that at the macroevolutionary level, such speciations 
will
necessarily appear stochastic and modelling these will necessitate 
relatively long computer simulations. Another reason is that, at this 
stage 
in the development of the subject, there are few guidelines on how to 
model 
speciation. It is clear that we need to go beyond population dynamics, 
and 
give some internal characteristics to each species which are then 
allowed 
to vary with time according to an adaptive dynamics. We have chosen 
discrete 
features as these internal characteristics, but a set of continuous 
features 
might have also been a viable choice. It would be interesting to 
explore 
other ways of defining what is meant by a species in evolutionary 
models 
of this type. We always assume that the time for an ecosystem to reach 
a
steady state is less than the time between speciation events. Thus it 
does
not matter if the speciation rate varies with the number of species, or 
has other similar factors influencing it. In the Webworld model, 
ecological
timescales are those which are longer than that defined by 
(\ref{popsize}), 
but shorter than the time between speciation events. On the other hand, 
evolutionary timescales are those which are longer (typically, much 
longer) 
than the time between speciation events.

There has been a discussion in the literature between those opposed to 
and 
those favouring a functional response, $g_{ij}$, which is 
ratio-dependent 
(Berryman, 1992; Huisman \& De Boer, 1997). Our main reason for 
choosing a 
ratio-dependent form was firstly, that it has many positive attributes 
from an ecological point of view (Berryman, 1992) and secondly, that it 
has 
fewer parameters associated with it. Indeed, our
generalization of the ratio-dependent functional response to a 
multispecies
food web (\ref{gij}), involves only two parameters: $b$ and $c$. 
Nevertheless, it is not clear whether the precise form of the 
functional 
response matters much in the context of Webworld, where the ultimate 
state of
interest is the one obtained after tens of thousands of speciation 
events.
It may be that only general aspects such as the existence of strong 
competition, particularly between similar species, and the inclusion of
a downward effect, as discussed earlier, are important. There may be a 
large degree of flexibility here, in the sense that the structure of 
the 
food web produced is insensitive to the precise form of $g_{ij}$, as 
long
as the right qualitative structure is present. Further work is needed 
to 
clarify these points and to identify what the important ingredients 
are.

The food webs generated by Webworld, some of which are shown in Figs. 
3-6,
seem to evolve in a ``natural'' way when they are examined time-step by
time-step. That is, the rules built into the model lead to consequences 
which can be interpreted according to rational criteria. Similarly, the 
consequences of a simple speciation event, which we have investigated 
in 
detail in this paper, seem very reasonable. Origination and extinction 
rates roughly balance on short time scales with slightly larger 
origination 
rates having an effect on longer time scales, child species tend to 
appear 
on the same level as parent species and there is strong competition 
between 
parent and child. All this is in line with expectations. 

In summary, Webworld is a model which covers time scales varying from 
the
very short, typified by changing foraging strategies, to the very long, 
required for evolutionary dynamics to reach a state where the number of 
originations and the number of extinctions balance on average. It gives 
results which are intuitively appealing and in broad agreement with 
food 
web data from real ecosystems. Many aspects of the model remain to be 
investigated, but we believe that it provides a realistic picture of 
the
evolution of ecological communities which throws light on the nature of 
the 
basic mechanisms present in all such communities.

\section*{Acknowledgment}

CQ wishes to thank EPSRC for the award of a postgraduate grant.
      
\newpage
\section*{References}

\noindent Arditi, R. \& Ginzburg, L.R. (1989). Coupling in
predator-prey dynamics: ratio-dependence. {\it J. Theor. Biol.}
{\bf 139}, 311-326.

\noindent Arditi, R. \& Michalski, J. (1995). Nonlinear food web
models and their responses to increased basal productivity. In:
{\it Food webs: integration of patterns and dynamics} (Polis, 
G.A. \& Winemiller, K.O., eds), pp.~122-133, Chapman \& Hall, London.

\noindent Bak, P. \& Sneppen, K. (1993). Punctuated equilibrium
and criticality in a simple model of evolution. {\it Phys. Rev.
Lett.} {\bf 71}, 4083-4086.

\noindent Bastolla, U., L\"assig, M., Manrubia, S.C. \& Valleriani, A.
(2000). Diversity patterns from ecological models at dynamical 
equilibrium. eprint: arXiv: nlin.AO/0009025.

\noindent Beddington, J. (1975). Mutual interference between parasites 
or 
predators and its effect on searching efficiency. {\it Anim. Ecol.} 
{\bf 51},
597-624.

\noindent Berryman, A.A. \& Millstein, J.A. (1989a). Are ecological
systems chaotic --- and if not, why not? {\it Trends Ecol. Evol.} {\bf 
4}, 
26-28.

\noindent Berryman, A.A. \& Millstein, J.A. (1989b). Avoiding 
chaos --- reply. {\it Trends Ecol. Evol.} {\bf 4}, 240-240.

\noindent Berryman, A.A. (1992). The origins and evolution of 
predator-prey theory. {\it Ecology} {\bf 73}, 1530-1535. 

\noindent Caldarelli, G., Higgs, P.G. \& McKane, A.J. (1998).
Modelling coevolution in multispecies communities. {\it J. Theor.
Biol.} {\bf 193}, 345-358.

\noindent Cohen, J.E. (1990). A stochastic theory of community
food webs VI - Heterogeneous alternatives to the cascade model.
{\it Theor. Pop. Biol.} {\bf 37}, 55-90.

\noindent Cohen, J.E., Briand, F. \& Newman, C.M. (1990). {\it
Biomathematics} Vol. 20. Community food webs, data and theory.
Springer Verlag, Berlin.

\noindent Drossel, B. (2001). Biological evolution and statistical 
physics. eprint: arXiv:cond-mat/0101409.

\noindent Drossel, B., Higgs, P.G. \& McKane, A.J. (2001). The
influence of predator-prey population dynamics on the long-term
evolution of food web structure. {\it J. Theor. Biol.} {\bf 208}, 
91-107.

\noindent Emlen, J.M. (1984). {\it Population biology}. Macmillan, 
New York.

\noindent Goldwasser, L. \& Roughgarden, J. (1993). Construction
and analysis of a large Caribbean food web. {\it Ecology} {\bf
74}, 1216-1233.

\noindent Hall, S.J. \& Raffaelli, D. (1991). Food web patterns:
lessons from a species-rich web. {\it J. Anim. Ecol.} {\bf 60},
823-842.

\noindent Hallam, T.G. (1986). Community dynamics in a homogeneous
environmemt.  In: {\it Mathematical ecology}. Biomathematics Vol. 17. 
(Hallam, T.G. and Levin, S.A., eds), pp.~241-285, Springer-Verlag, 
Berlin.

\noindent Hastings, A. \& Powell, T. (1991). Chaos in a three-species 
food chain, {\it Ecology} {\bf 72}, 896-903.

\noindent Hofbauer, J. \& Sigmund, K. (1998). {\it Evolutionary games 
and 
population dynamics}. Cambridge University Press, Cambridge.

\noindent Holling, C.S. (1959). Some characteristics of simple types of
predation and parasitism. {\it Can. Entomol.} {\bf 91}, 385-398.

\noindent Huisman, G. \& De Boer, R.J. (1997). A formal derivation 
of the Beddington functional response, {\it J. Theor. Biol.} {\bf 185}, 
389-400, and references therein.

\noindent L\"assig, M., Bastolla, U., Manrubia, S.C. \& Valleriani, A.
(2001). The shape of ecological networks. eprint: 
arXiv:nlin.AO/0101026.

\noindent Logofet, D.O. (1993). {\it Matrices and graphs: stability 
problems in mathematical ecology}. CRC Press, London.

\noindent Martinez, N.D. \& Lawton, J.H. (1995). Scale and food
web structure --- from local to global. {\it Oikos} {\bf 73},
148-154.

\noindent May, R.M. (1974). {\it Stability and complexity in model 
ecosystems}. Monographs in population biology, Vol. 6. Princeton 
University 
Press, Princeton. Second edition.

\noindent McCann, K. \& Yodzis, P. (1994). Biological conditions for
chaos in three-species food chain. {\it Ecology} {\bf 75}, 561-564.

\noindent Parker, G.A. \& Maynard Smith, J. (1990). Optimality
theory in evolutionary biology. {\it Nature}. {\bf 348}, 27-33.

\noindent Pielou, E.C. (1977). {\it Mathematical ecology}. Wiley, 
New York, Second edition.

\noindent Pimm, S.L. (1982). {\it Food webs}. Chapman \& Hall,
London.

\noindent Post, D.M., Conners, M.E. \& Goldberg, D.S. (2000). Prey 
preference by a top predator and the stability of linked food chains. 
{\it Ecology} {\bf 81}, 8-14, and references therein.

\noindent Reeve, H.K \& Dugatkin, L.A. (1998). Why we need
evolutionary game theory. In: {\it Game theory and animal behaviour}
(Dugatkin, L.A \& Reeve, H.K., eds), pp.~304-311. Oxford University 
Press, Oxford.

\noindent Roughgarden, J. (1979). {\it Theory of population genetics 
and
evolutionary ecology}. Macmillan, New York.

\noindent Sol\'e, R.V., Manrubia, S.C., Benton, M. \& Bak, P.
(1997). Self-similarity of extinction statistics in the fossil
record. {\it Nature} {\bf 388}, 764-767.

\noindent Stephens, D. W. \& Krebs, J. R. (1986). {\it Foraging 
theory}. Princeton University Press, NJ. 

\noindent Svirezhev, Yu.M. \& Logofet, D.O. (1983). {\it Stability of 
biological communities}. Mir Publishers, Moscow.

\end{document}